\documentclass[%
 reprint, 
showpacs,
 amsmath,amssymb,
 aps,
prb,
floatfix,
 lengthcheck,
]{revtex4-1}

\usepackage{graphicx}

\begin{document}


\renewcommand\Re{\operatorname{Re}}
\renewcommand\Im{\operatorname{Im}}
\newcommand{\Tr}{\operatorname{Tr}}

\newcommand{\HOLU}{{\sc HOMO-LUMO} }
\newcommand{\LU}{{\sc LUMO} }
\newcommand{\HO}{{\sc HOMO} }

\title{Coherent Destruction of Coulomb Blockade Peaks in Molecular Junctions} 

\author{J.~P.~Bergfield}
\affiliation{College of Optical Sciences, University of Arizona, 1630 East University Boulevard, AZ 85721}
\author{Ph.~Jacquod}
\affiliation{Department of Physics, University of Arizona, 1118 East Fourth Street, Tucson, AZ 85721}
\author{C.~A.~Stafford}
\affiliation{Department of Physics, University of Arizona, 1118 East Fourth Street, Tucson, AZ 85721}

\pacs{
73.23.Hk 	
85.35.Ds 	
73.63.-b, 
85.65.+h, 
}

\date{\today}

\begin{abstract}
Coherent electronic transport in single-molecule junctions is investigated in the Coulomb blockade regime.  Both the transmission phase and probability are calculated for junctions with various contact symmetries.  A dramatic suppression of the Coulomb blockade peaks is predicted for junctions where multiple atomic orbitals of the molecule couple to a single electrode although the charging steps are unaffected. 
\end{abstract}

\maketitle

\section{Introduction}
Coulomb blockade has been investigated in mesoscopic transport through metallic nanoparticles~\cite{Delft01} and semiconductor quantum dots,\cite{Alhassid00,Aleiner02} and more recently in molecular transport.\cite{Kubatkin03,Poot06,Bergfield09,Song09} Because the distance between conductance peaks is set by the charging energy, one might naively think that transport in the Coulomb blockade regime is essentially incoherent.  Recent interferometric experiments have shown, however, that electrons transferred through quantum dots retain memory of their phase,\cite{Schuster97,Avinun-Kalish05} a finding that has triggered intensive theoretical investigations on the transmission phase through interacting systems.\cite{Lee99,LevyYayati00,Silverstrov00,Hackenbroich01,Silva02,Karrasch07}


Mesoscopic experiments access regimes where tunneling transport occurs either between two discrete channels (in lateral quantum dots) or two quasi-continua (in metallic nanoparticles) that are connected by a central quantum system with no particular symmetry.  Consequently, interchannel coherence is either absent or washed out, and the spatial symmetry with which external terminals are connected to the central system matters little.  The situation is fundamentally different in molecular transport, because molecules naturally possess symmetries that strongly affect the phase coherent transmission of electrons.\cite{Cardamone06}  Moreover, molecular structure makes it possible to tunnel-couple multiple orbitals to a single external electrode.\cite{Kiguchi08}

\begin{figure}	
\includegraphics[width=\linewidth]{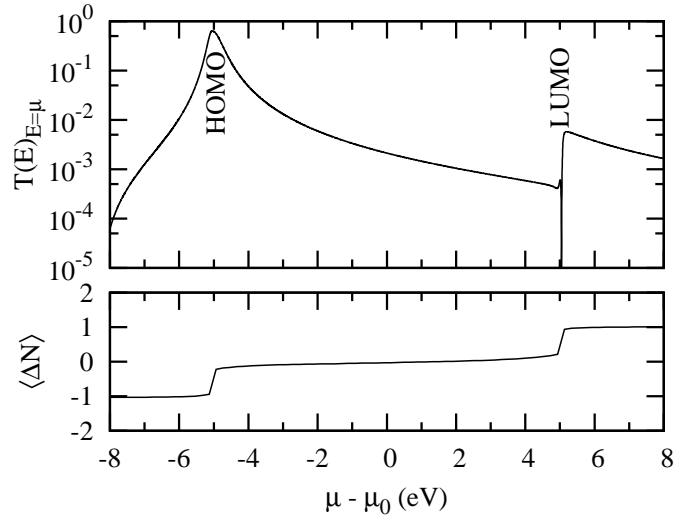}
\caption{Transmission spectrum and average electron occupancy of a 
flat ethylene (C$_2$H$_2$) molecular junction in which one lead couples symmetrically to 
both $\pi$-orbitals of the molecule and the other couples only to one orbital. 
The HOMO resonance of the molecule yields a typical transmission resonance, but transmission is blocked completely at the LUMO resonance 
by destructive quantum interference.  Despite the absence of a LUMO transmission resonance, the HOMO and LUMO charging steps are virtually identical.
%
%
%
}
	\label{fig:diatomic_figure1}
\end{figure}


In this article, we consider junctions composed of small molecules with $n$-fold spatial symmetry
coupled to two external single-channel leads.  The molecule-lead coupling can occur either via single or multiple spatially separated atomic orbitals.  For single-orbital coupling we show that the presence of nodes in the junction's transmission spectrum depends critically upon the contact geometry of the junction.
This is so, because the connectivity determines the relative phase accumulated by an electron along
different, interfering paths across the molecule. 

The presence of such a node results in an additional $\pi$ shift in the junction's transmission phase.\cite{Lee99,LevyYayati00} This is of crucial importance for multi-orbital coupling, because this phase shift can turn constructive interferences into destructive ones, thereby suppressing the conductance.  
Accordingly, we find that Coulomb blockade resonances can be completely suppressed in molecular junctions where several atomic orbitals of the molecule 
couple to a single electrode.

\begin{figure*}[tbh]
\begin{picture}(0,0)
\put(40, 212){\includegraphics[width=.8in]{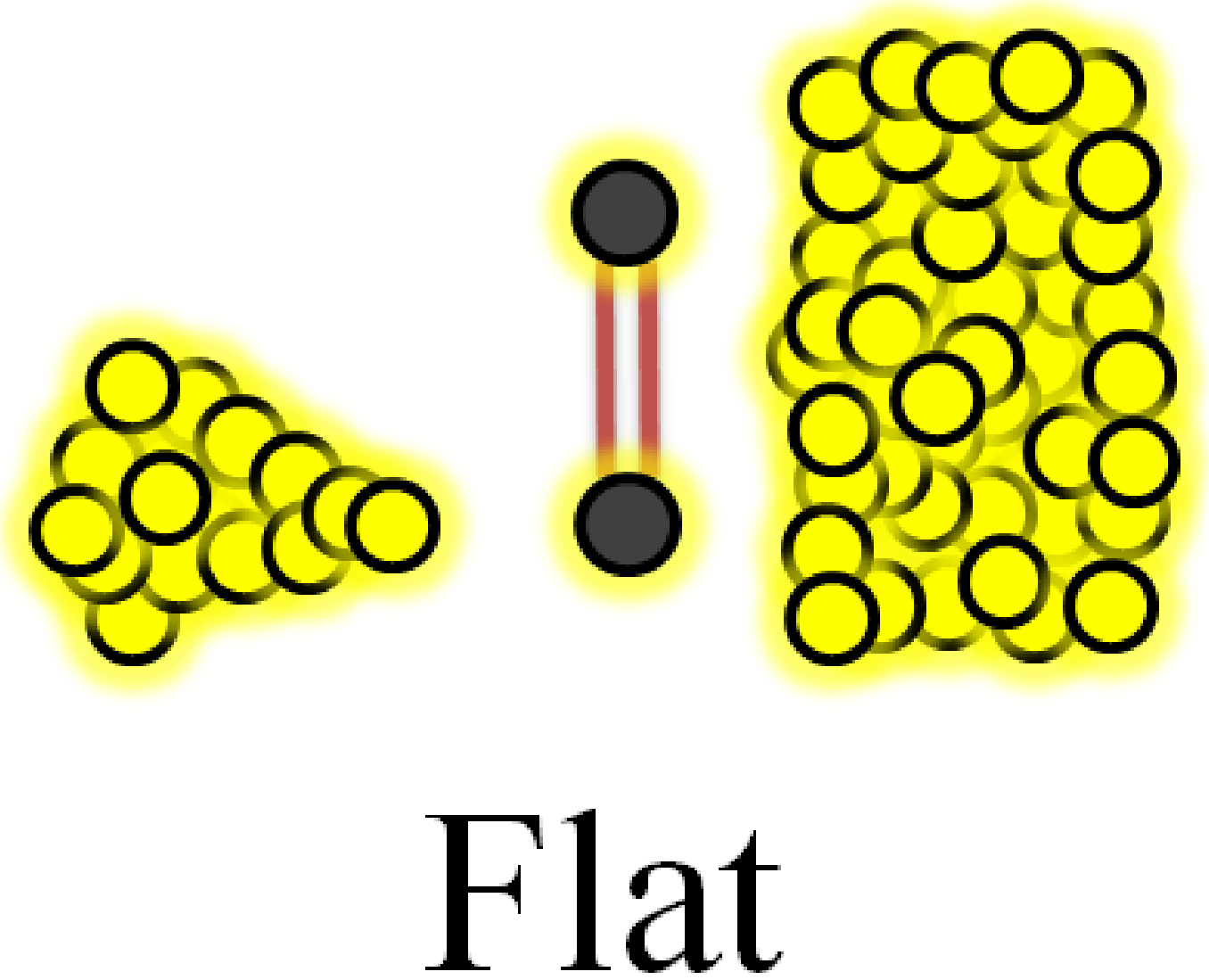}}
\put(115, 212){\includegraphics[width=.9in]{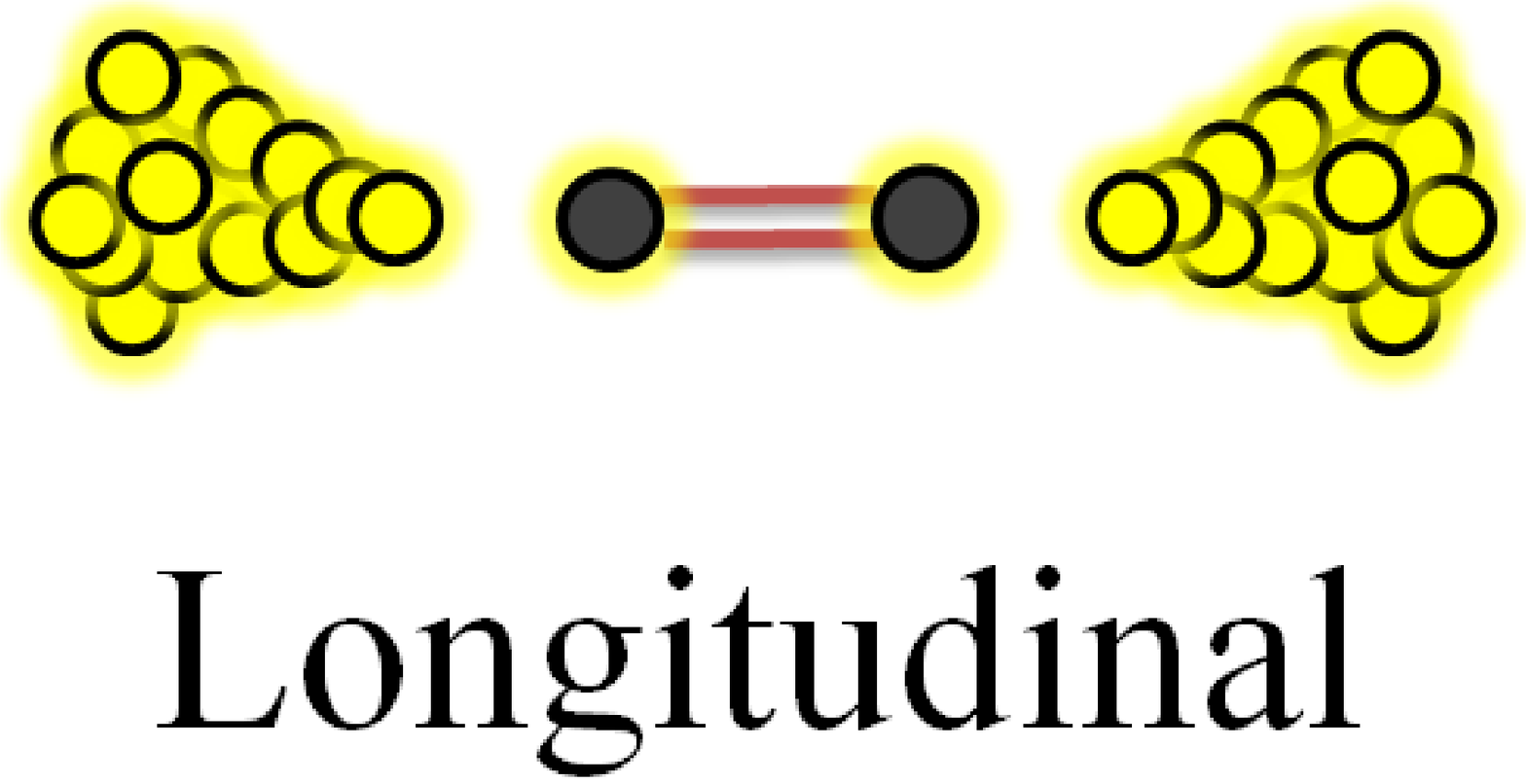}}
\put(200, 212){\includegraphics[width=.8in]{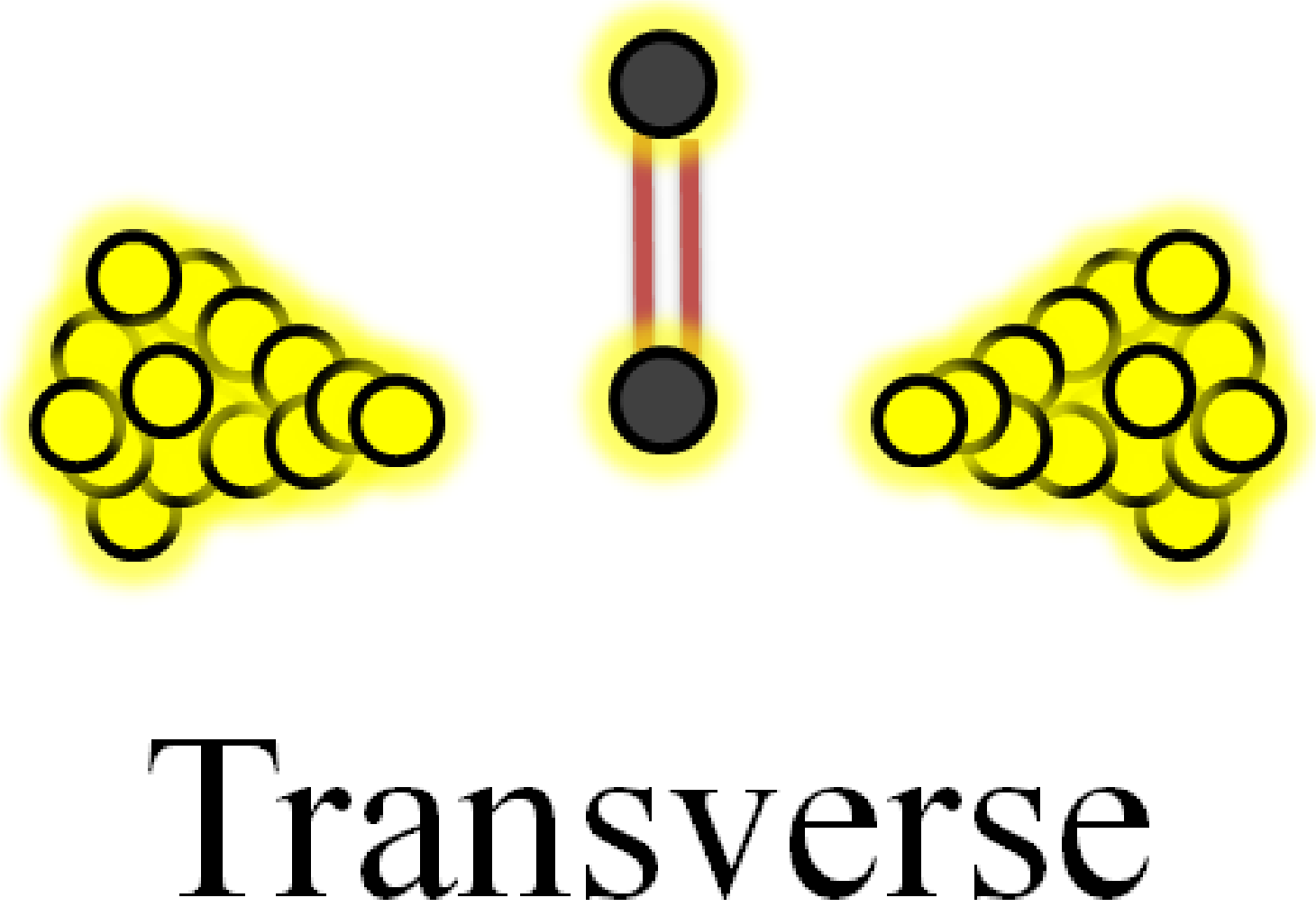}}
\end{picture}
	\centering
		\includegraphics[width=.99\linewidth]{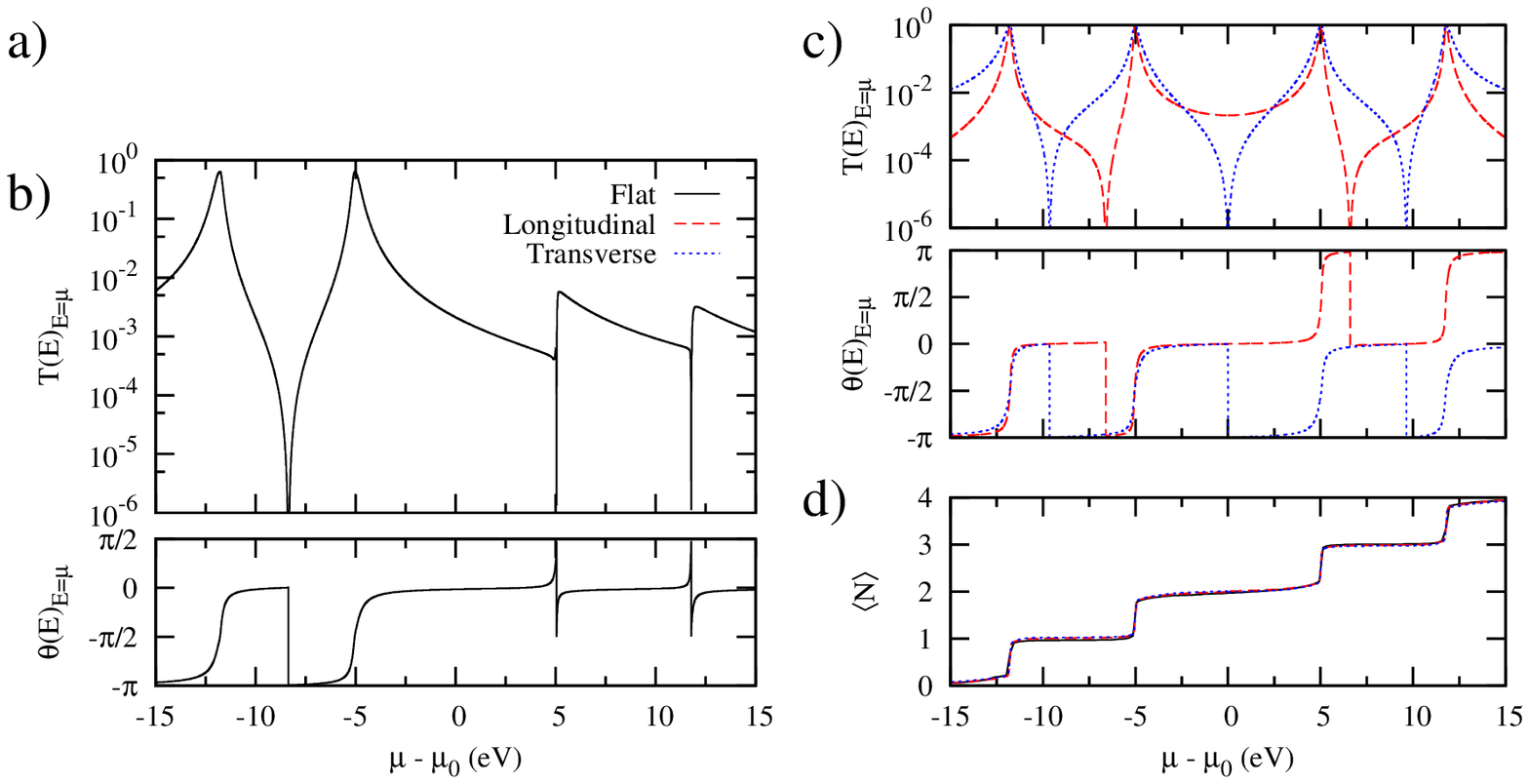}
	\caption{Numerically obtained transport results for flat, longitudinal and transverse ethylene (C$_2$H$_4$) junctions. (a) Schematic diagrams for each junction's geometry.  (b) and (c) Transmission probabilities (top) and phases (bottom) for the flat (b) and transverse and longitudinal (c) connections.  (d) Molecular $\pi$-orbital occupancy for all three geometries, showing that the destruction of the Coulomb peaks does not affect the accumulation of charge on the molecule.  
The transmission spectrum in both the longitudinal and transverse geometries is symmetric about $\mu_0$, whereas it is asymmetric in the flat configuration, a property which should be experimentally resolvable. 
	Using the molecular Hamiltonian and parameterization of Ref.~[\onlinecite{Bergfield09}], the parameters for ethylene were: $t$=2.6eV, U$_{11}$=8.9eV, U$_{12}$=5.33eV and $\Gamma$=$\Gamma_{\rm I}$=$\Gamma_{\rm II}$=0.5eV in the longitudinal and transverse geometries while $\left[ \gamma_{\rm II} \right]_{nm}$=$\sqrt{\Gamma}$/2 in the flat geometry.
	}
	\label{fig:diatomic_flat_vs_longitudinal}
\end{figure*}

This is the principal message of this article, and is illustrated in Fig.~\ref{fig:diatomic_figure1}, where the transmission spectrum and average 
electronic occupancy of a flat 
ethylene (C$_2$H$_2$) junction is plotted against the electrochemical potential $\mu$.  
In the flat junction, one lead couples symmetrically to
both $\pi$-orbitals of the molecule while the other couples only to one orbital (see Fig.\ \ref{fig:diatomic_flat_vs_longitudinal}a).
This junction geometry can be experimentally realized for many molecular species if the molecule is
deposited on a metallic surface and contacted from above by an STM tip\cite{Repp05,Repp06} or in a mechanically controlled break junction.\cite{Kiguchi08}  
In Fig.~\ref{fig:diatomic_figure1}, 
the Highest Occupied Molecular Orbital (HOMO) resonance of the molecule yields a typical transmission resonance 
but transmission is blocked completely at the Lowest Unoccupied Molecular Orbital (LUMO) resonance
by destructive quantum interference, leading to a transmission node.  
Destructive interference is not complete in the tails of the LUMO resonance, but the transmission is still suppressed by two orders of magnitude.
Despite the absence of a LUMO transmission resonance, the LUMO charging step illustrated in the lower panel of Fig.~\ref{fig:diatomic_figure1}
is essentially the same as that at the HOMO resonance.  Thus we have the {\em phenomenon of Coulomb blockade without a Coulomb blockade peak}.

This paper is organized as follows.  In Sec.\ \ref{sec:transport_formalism} we outline the transport formalism necessary to describe single-molecule junctions with multi-orbital lead-molecule coupling.  The linear-response transport of ethylene and benzene-based junctions are investigated for several junction bonding configurations in Sec.\ \ref{sec:linear_response}.  The non-linear transport of flat and para benzene junctions are presented in Sec.\ \ref{sec:non_linear_response}.  Conclusions and discussion are presented in Sec.\ \ref{sec:conclusions}.

\section{Transport formalism}
\label{sec:transport_formalism}
Transport quantities are naturally expressed in terms of Green's functions.  The retarded Green's function $G_{n\sigma,m\sigma'}(t)= -i\theta(t)\langle \{d_{n\sigma}(t),d_{m\sigma'}^\dagger(0)\}\rangle$ of the junction may be transformed into the energy domain giving:\cite{Bergfield09}
\begin{equation}
G(E) = \left[ {\bf 1}E - H_{\rm mol}^{(1)} - \Sigma_{\rm T}(E) - \Sigma_{\rm C}(E)\right]^{-1},
\label{eq:Dyson2}
\end{equation}
where $H_{\rm mol}^{(1)}$ is the one-body part of the molecular Hamiltonian $H_{\rm mol}$, $\Sigma_{\rm T}$ is the tunneling self-energy matrix and $\Sigma_{\rm C}$ is the Coulomb self-energy matrix.  $H_{\rm mol}$ is modeled using a semi-empirical Hamiltonian known to accurately characterize the electronic spectrum of $\pi$-conjugated molecules.\cite{Bergfield09,Castleton02}  The tunneling self-energy $\Sigma_{\rm T}$=$\sum_{\alpha}\Sigma^\alpha_{\rm T}$ can be decomposed as the sum of contributions from the various contacts, which can be calculated exactly using an equations-of-motion method.\cite{Bergfield09, Jauho94}  Of particular interest is the imaginary part of $\Sigma_{\rm T}$, which causes broadening of the molecular resonances.  Often this broadening is discussed in terms of the tunneling-width matrix, defined for lead $\alpha$ as:\cite{Jauho94}
\begin{equation}
\Gamma^\alpha(E)=i\left(\Sigma_{\rm T}^\alpha(E) -\left[\Sigma_{\rm T}^{\alpha}(E)\right]^\dagger \right).
\label{eq:DefOfGamma}   
\end{equation}
The Coulomb self-energy must in general be calculated using an appropriate approximation.  Here we utilize the non-perturbative method of Ref.~[\onlinecite{Bergfield09}] to calculate $\Sigma_{\rm C}$, which has been shown to accurately describe Coulomb blockade in single-molecule junctions.


While the Green's function fully characterizes the molecular junction, the transmission phase is most readily defined in terms of the scattering matrix $S_{\alpha\beta}(E)$.  For purely elastic quantum transport, the dominant transport mechanism at room temperature in single-molecule junctions, the Green's function and scattering matrix are related as follows:\cite{Fischer81,LevyYayati00}
\begin{equation}
S_{\alpha\beta}(E) = -\delta_{\alpha\beta}{\bf 1} + i \gamma_{\alpha}(E)G(E)\gamma_{\beta}^\dagger(E).
	\label{eq:fischer_lee}
\end{equation}
The off-diagonal terms of $S$ are the transmission amplitudes $\left[t_{\alpha\beta}(E)\right]_{nm}$ for scattering from mode $m$ in lead $\beta$ to mode $n$ in lead $\alpha$.  The total transmission probability between leads $\alpha$ and $\beta$ is given by:
\begin{equation}
T_{\alpha \beta}(E) = \Tr \left\{t_{\alpha\beta}(E) t_{\alpha\beta}^\dagger(E) \right\},	
\label{eq:Tab}
\end{equation}
where the trace is over all relevant atomic orbitals of the molecule.  The tunneling-width amplitude matrices $\gamma_\alpha(E)$ are related to the tunneling-width matrices $\Gamma^\alpha(E)$ as follows:
\begin{equation}
 \Gamma^{\alpha}(E)=\gamma_\alpha^\dagger(E) \gamma_{\alpha} (E).
 \label{eq:Gamma}
 \end{equation}

In order to simplify our present discussion and analysis we shall henceforth: (i) Consider only two-terminal junctions, (ii) assume each lead is characterized by a single mode, and (iii) take the broad-band limit~\cite{Jauho94} for the lead-molecule coupling, where $\Gamma^\alpha(E)$$\equiv$$\Gamma^\alpha$ becomes an energy-independent matrix of constants related to the tunneling rate.  Together, the first two assumptions ensure the existence of a single well-defined transmission eigenphase $\theta(E)$=$\arg\left[t(E)\right]$ for the junction.  


%

Although the investigation of coherent quantum transport in molecular junctions is not new,\cite{Nitzan03} most theoretical studies limit their analysis to cases where $\Gamma^\alpha$ is diagonal, corresponding physically to the special case where single atomic orbitals couple to single leads.  Presently, we consider transport in a junction geometry we label `flat,' where a single orbital connects to one lead while all atomic orbitals of the molecule couple to a single channel (e.g., $s$-wave) of the surface electronic states in the other, a configuration which is already experimentally accessible.\cite{Repp05,Repp06} 

%

\section{Linear-response junction transport}
\label{sec:linear_response}


In a flat junction with two leads, labeled I and II, the tunneling-amplitude matrices are $\left[ \gamma_{\rm I} \right]_{nm}$=$\sqrt{\Gamma_{\rm I}} \delta_{na}\delta_{ma}$ and $\left[ \gamma_{\rm II} \right]_{nm}$=$\sqrt{\Gamma_{\rm II}}$, where $a$ is the orbital connected to lead I and all orbitals couple equally to lead II.  $\Gamma_{\rm I}$ and $\Gamma_{\rm II}$ are constants characterizing the tunneling between the molecule and leads I and II, respectively.  Following Eq.~(\ref{eq:fischer_lee}), the transmission amplitude becomes: 
\begin{equation}
t_{\rm I,II}^{\rm Flat}(E)=i\sqrt{\Gamma_{\rm I}\Gamma_{\rm II}} \sum_{j=1}^{N}G_{aj}(E),
\label{eq:tLR_flat}
\end{equation}
where $N$ is the total number of molecular orbitals.  From Eq.~(\ref{eq:Tab}), we see that the transmission probability will be composed of diagonal terms proportional to $\sum_{j}|G_{aj}|^2$ and cross terms proportional to $\sum_{j \neq k}G_{aj}G^*_{ak}$.  We now show that interferences between various transmission amplitudes can cause these two terms to cancel, giving rise to a dramatic reduction of the transmission peaks. 

\subsection{Ethylene junction}
For pedagogical reasons, we first consider several ethylene (C$_2$H$_4$) junctions, each shown schematically in Fig.~\ref{fig:diatomic_flat_vs_longitudinal}a.  The transmission probability and phase for the flat junction is shown in Fig.~\ref{fig:diatomic_flat_vs_longitudinal}b, while the spectra for the transverse and longitudinal junctions are both shown in Fig.~\ref{fig:diatomic_flat_vs_longitudinal}c.  In all three geometries $\Gamma_{\rm I}$=0.5eV.  In the longitudinal and transverse cases $\Gamma_{\rm II}$=0.5eV, while in the flat geometry $\Gamma_{\rm II}$=0.25eV such that the total coupling $\tilde{\Gamma}_{\rm II}$=$\Tr \left\{ \Gamma^{\rm II} \right\}$=0.5eV.  

For chemical potentials below the center of the HOMO-LUMO gap ($\mu_0$), the spectra of the three geometries are qualitatively very similar: for every peak in one there is a peak in the others and similarly for the nodes, although the exact location of the nodes varies.  For $\mu>\mu_0$, however, there is a nearly 500$\times$ reduction of the Coulomb blockade peak height in the flat geometry as compared with either the longitudinal or transverse spectra. 

The origin of this suppression can be understood by examining the transmission amplitudes in each geometry.  In the longitudinal and transverse junctions we find $t_{\rm I,II}^{\rm Long}(E)$=$i\Gamma \left[G_{12}(E)\right]$ and $t_{\rm I,II}^{\rm Trans}(E)$=$i\Gamma \left[ G_{11}(E)\right]$, respectively, where $\Gamma_{\rm I}$=$\Gamma_{\rm II}$=$\Gamma$.  From Eq.~(\ref{eq:tLR_flat}), we see that the flat junction's amplitude $t_{\rm I,II}^{\rm Flat}(E)$=$i\Gamma/\sqrt{2} \left[ G_{11}(E)+ G_{12}(E)\right]$  is proportional to the sum of the amplitudes from the other two configurations.  From the longitudinal and transverse phase spectra, shown in the lower part of Fig.~\ref{fig:diatomic_flat_vs_longitudinal}c, it is evident that the node in $G_{11}$ when $\mu$=$\mu_0$ causes $G_{11}$ and $G_{12}$ to become $\pi$ out of phase for all $\mu >\mu_0$.  Exactly on resonance with $\mu > \mu_0$, $|G_{11}|$=$|G_{12}|$ and consequently, via total destructive interference, a transmission {\em peak} in the longitudinal (or transverse) spectrum becomes a transmission {\em node} in the flat junction's spectrum, with a concomitant lapse of the transmission phase in the latter.  In the vicinity of a resonance, the two terms are not exactly equal but still interfere destructively giving rise to a pronounced reduction of the tails of the Coulomb blockade peaks instead of an exact cancellation.

Transmission peaks occur at energies where a system has a degeneracy between two charge states.  Even though these Coulomb blockade peaks may be nearly destroyed by quantum interferences, as we just showed, the charge of the molecule  still changes by one as we cross each resonance.  The $\pi$-orbital molecular occupancy is shown for all three junction geometries as a function of chemical potential in Fig.~\ref{fig:diatomic_flat_vs_longitudinal}d and exhibits nearly the same spectra for all three geometries.  Comparing the phase of transmission, shown in the bottom parts of Figs.~(\ref{fig:diatomic_flat_vs_longitudinal}b,\ref{fig:diatomic_flat_vs_longitudinal}c) with the charge spectra, we see that for each step in the occupancy there is an associated increase in the transmission phase by $\pi$, in agreement with the Friedel-sum rule.\cite{Friedel58}

\subsection{Benzene junction}
\label{sec:benzene_junction}
Since the suppression of transmission peaks in flat junctions are a manifestation of interference between multiple transport pathways, the effect should be more pronounced in larger molecules which possess a correspondingly larger number of transport pathways. To investigate this hypothesis, we consider the transmission spectrum for a flat benzene junction in which one electrode couples to just a single $\pi$-orbital of the benzene ring, while the second couples equally to all six $\pi$-orbitals.  Such a junction could be experimentally realized by depositing a benzene molecule on a Pt or graphene surface and contacting it from above with an STM tip\cite{Repp05,Repp06} or using a mechanically controlled break junction.\cite{Kiguchi08} 

\begin{figure}[tb]
	\centering
	\begin{picture}(0,0)
\put(-55, -35){\includegraphics[width=1.0in]{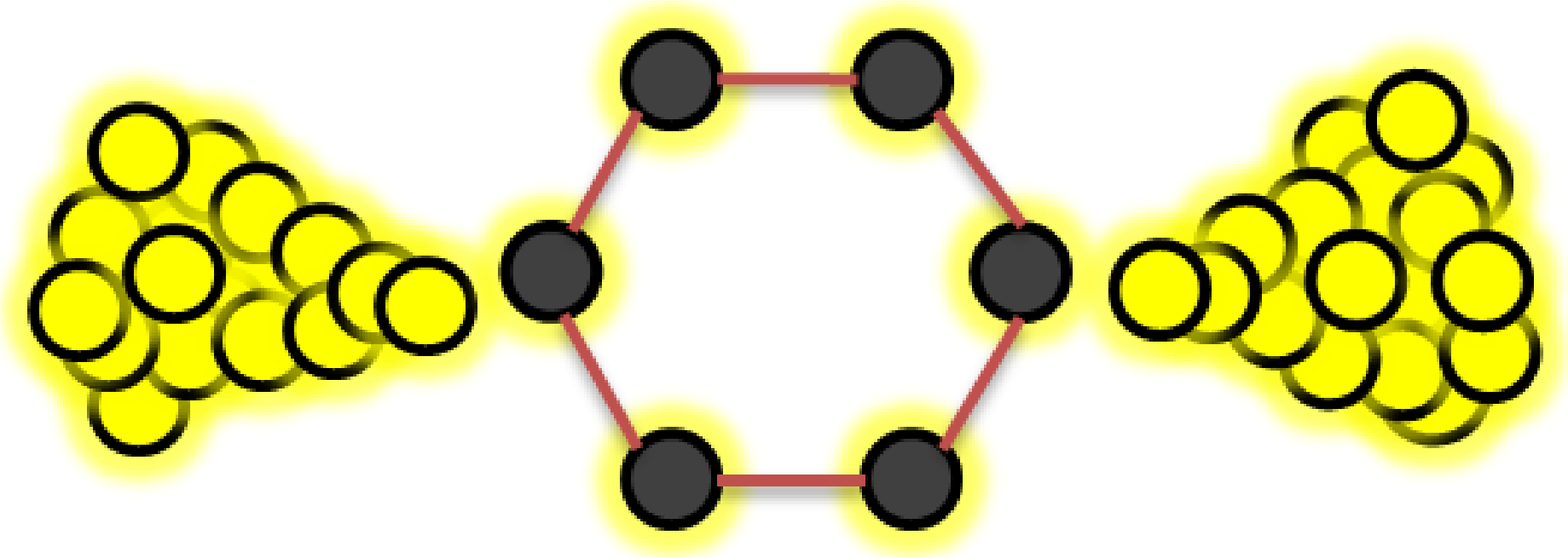}}
\put(70, -43){\includegraphics[width=.5in]{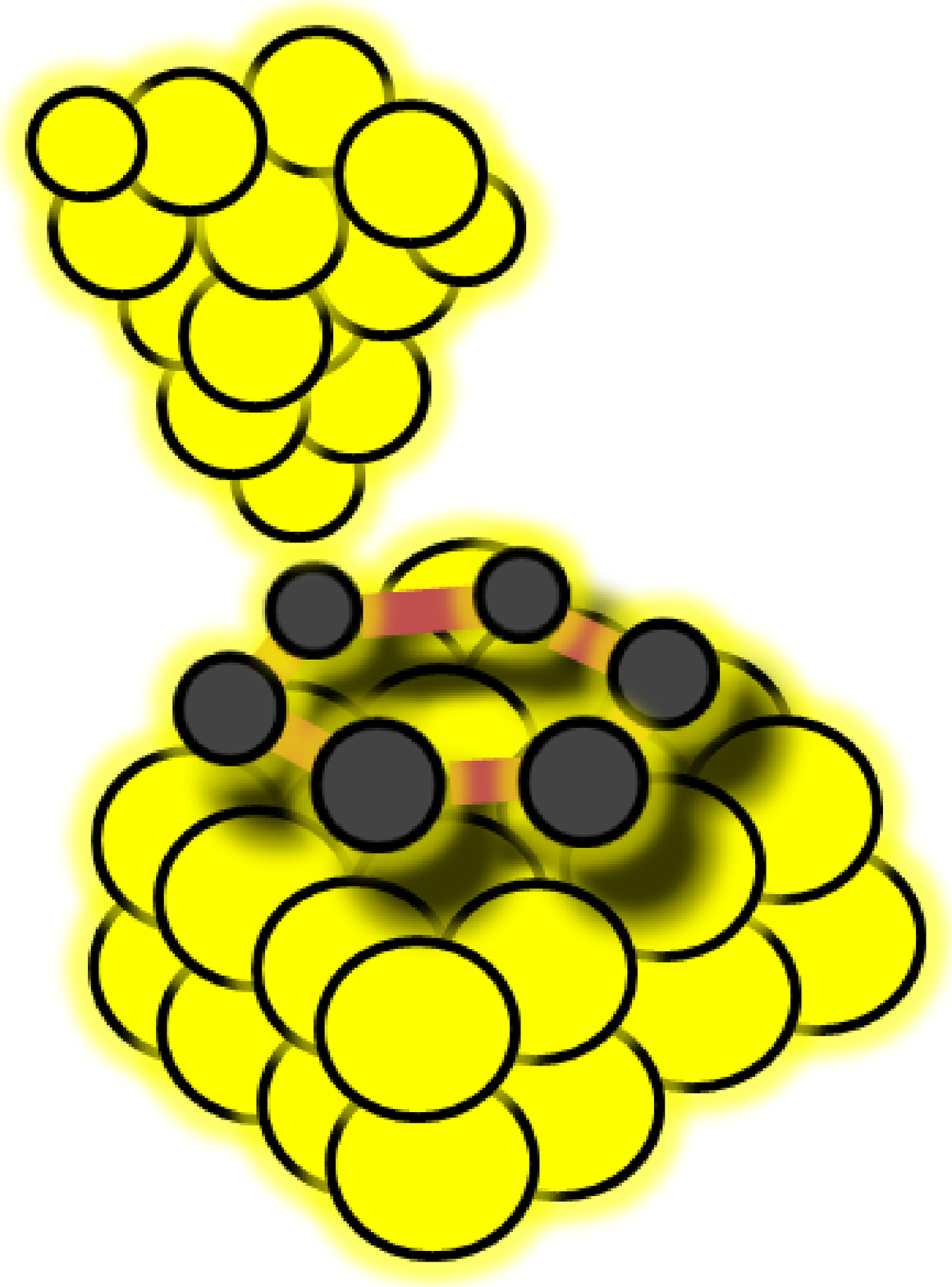}}
\end{picture}	
	\includegraphics[width=\linewidth]{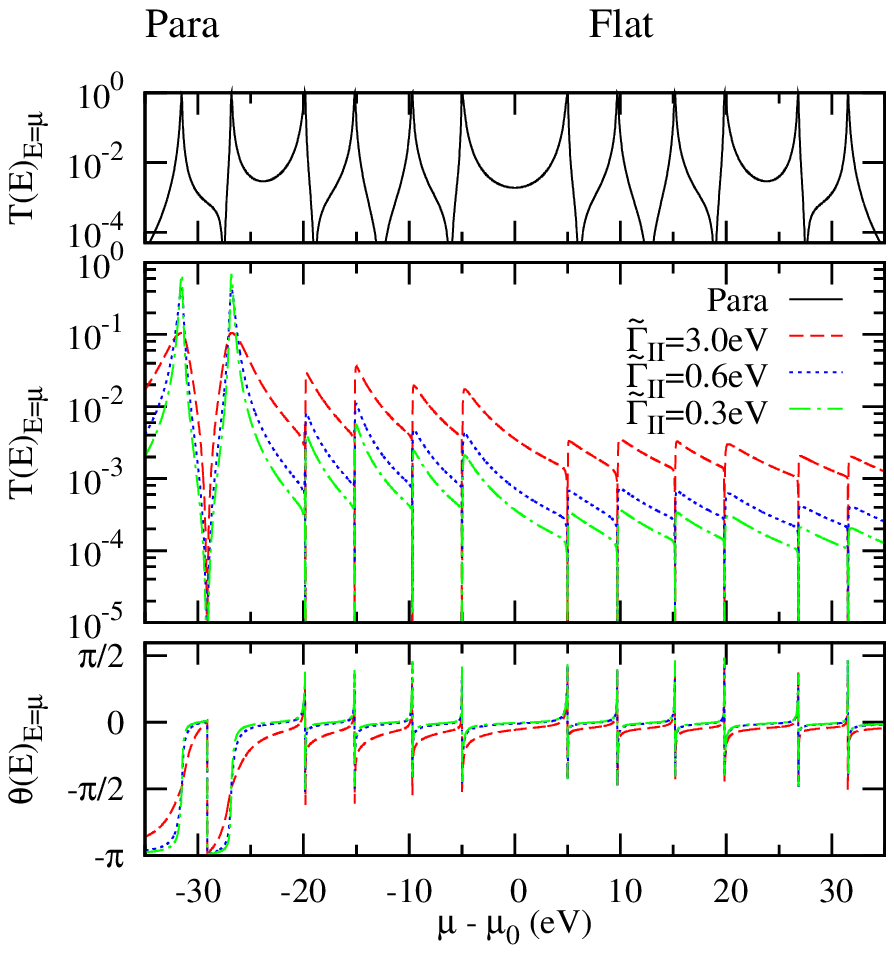}
	\caption[benzene]{
	Top panel: Sketch of the para and the flat, multiorbital
coupling geometries for a benzene (C$_6$H$_6$) molecular
junction.  Second panel: Transmission spectrum (logscale) of a para benzene junction.  Lowest two panels: Transmission probability and phase for a flat benzene junction.  All but the first two transmission peaks are completely suppressed in the
flat junction due to destructive quantum interference.  The destructive interference is not complete in the tails of the resonances, which 
exhibit a maximal $\sim$3350$\times$ reduction for $\tilde{\Gamma}_{\rm II}$=0.3eV.  Reductions of $\sim$1050$\times$ and $\sim$50$\times$ were found for $\tilde{\Gamma}_{\rm II}$=0.6eV and $\tilde{\Gamma}_{\rm II}$=3.0eV, respectively.  
The full transmission spectrum of the para benzene junction, shown in the top portion of this figure, is symmetric about $\mu_0$ while that of the flat 
junction is strongly asymmetric.  
The complete spectrum of each junction is included for illustrative purposes, although the full range of gating may not be experimentally achievable.  In all cases, lead I is coupled to a single orbital with $\Gamma_{\rm I}$=0.6eV.  In the para junction $\Gamma_{\rm II}$=0.6eV.  The total coupling to the flat-contacted second lead $\tilde{\Gamma}_{\rm II}$ was varied, where each orbital contributed $\tilde{\Gamma}_{\rm II}$/6.  The model and parameterization used in these simulations are discussed in detail in Ref.~[\onlinecite{Bergfield09}].
}
\label{fig:benzene_flat_vs_para}
\end{figure}

The calculated spectrum of a flat benzene junction is shown in Fig.~\ref{fig:benzene_flat_vs_para} for several values of lead-molecule coupling.  Except for the first two peaks, all the molecular resonances coincide with transmission nodes, a fact which is observable from the transmission probability or as $\pi$-slips in the transmission phase, shown in the bottom portion of the same figure.  Owing to the molecular symmetry of benzene, there are only four non-degenerate terms in the flat junction's transmission amplitude.  Using Eq.~(\ref{eq:tLR_flat}):
\begin{equation}
	t_{\rm I,II}^{\rm Flat} = i\sqrt{\Gamma_{\rm I}\Gamma_{\rm II}} \left[ t_{11} + \underbrace{t_{14}}_{\rm para} + 2\underbrace{t_{12}}_{\rm ortho} + 2 \underbrace{t_{13}}_{\rm meta} \right],
	\label{eq:BenzeneTFlat}
\end{equation}
where the {\em ortho} and {\em meta} terms are doubly degenerate and the dependence on energy is implicit.  The first two peaks correspond to resonant tunneling through the nodeless molecular ground state, for which no destructive interference is possible, so that all the terms in Eq.~(\ref{eq:BenzeneTFlat}) add constructively.  For every other molecular resonance, three amplitudes are in phase and three are $\pi$ out of phase such that by symmetry the total transmission amplitude vanishes exactly on resonance.  Elsewhere in the spectrum, this {\em coordination of cancellation} is visible as a strong suppression of the molecular resonance peak heights.  As the lead-molecule coupling is reduced the individual transmission amplitudes are only appreciable near a resonance, where destructive interference between amplitudes is most complete, resulting in enhanced peak suppression with {\em decreasing} lead-molecule coupling.

\section{Non-linear junction transport}
\label{sec:non_linear_response}

As we have already seen, the transmission probability of the junction, which determines the linear-response transport coefficients, is strongly affected by the contact geometry.  In the following discussion, we focus on the non-linear junction response of a flat junction composed of a single benzene molecule adsorbed on a Cu(110) surface.  Although there are studies for other Cu surfaces,\cite{Lauhon00,Weiss98,Rogers04,Bilic06,Lesnard08} we focus on the 110 plane because low-coverage experiments have already been performed for benzene using a STM setup similar to what we propose.\cite{Komeda04}

%

\begin{figure}[tb]
	\centering	
\includegraphics[width=\linewidth]{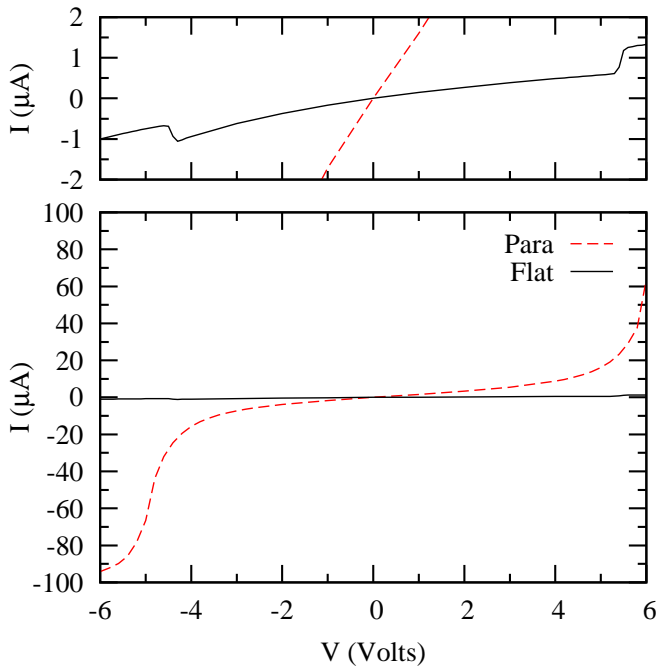}
	\caption{The non-linear current $I$ as a function of bias voltage $V$ for para and flat benzene-Cu junctions.  Because of the coherent suppression of the Coulomb blockade peaks in the flat geometry, features in the current as a function of voltage are strongly reduced as compared to the para geometry.  A closeup of the current in the flat junction is shown in the top portion of the figure.  In the flat junction, the benzene molecule lies 2.7{\AA} above the Cu(110) plane and 5{\AA} below the STM tip with lead-molecule coupling matrices $\Gamma^{\rm II}_{nm}$=1.3eV and $\Gamma^{\rm I}_{nm}$=$\delta_{na}\delta_{ma}$0.1eV, respectively, where $a$ is an atomic orbital of the molecule.  In the para junction both the copper surface and STM tip lie 3{\AA} from their closest respective atomic orbitals and each couples with 1.3eV to that orbital.  The voltage $V$=$V_{\rm I}$-$V_{\rm II}$, where $V_{\rm I}$ is the STM voltage and $V_{\rm II}$ is the voltage of the Cu(110) surface.}
	\label{fig:flat_benzene_IV}
\end{figure}


Experimentally, it has been determined that $\Delta E_{\rm b}$$\leq$1.0eV (99kJ/mol), where
$\Delta E_{\rm b}$ is the binding energy of benzene on Cu(110) surface.\cite{Rogers04}
Using many-body perturbation theory the binding energy can be related to the lead-molecule coupling\cite{BergfieldUnpublished} $\Gamma$ where, for a flat configuration in which the benzene molecule lies 2.7{\AA} above the copper surface, we find $\Tr\{\Gamma\}\leq$7.8eV, {ca.}\ 1.3eV per atomic orbital of the molecule.  Screening of the intramolecular Coulomb interactions by the copper surface was included via the image charge method using a semi-empirical multipole expansion for the $\pi$-electrons.\cite{JoshUnpublished}



%
%
%


The simulated non-linear responses of the para and flat junctions are shown in Fig.\ (\ref{fig:flat_benzene_IV}), where the current flowing into lead $\alpha$ for a two-terminal device in the elastic cotunelling regime\cite{Bergfield09} is given by:\cite{Buttiker86} 
\begin{equation}
	I_\alpha = \int_{-\infty}^{\infty} dE \; T_{\alpha\beta}(E) \left[f_\beta(E) - f_\alpha(E) \right],
	\label{eq:current}
\end{equation}
where $f_\alpha(E)$ is the Fermi distribution for lead $\alpha$ at chemical potential $\mu_\alpha$ and inverse temperature $\beta_\alpha$.  Relative to the center of the \HOLU gap $\mu_{\rm Cu(110)}$-$\mu_0$=-0.425eV, where we have used the work function of the Cu(110) surface\cite{CRC} $\phi_{\rm Cu(110)}$=4.48eV and $\mu_0$=$(\varepsilon_{\rm IE}+\varepsilon_{\rm EA})/2$=-4.055eV.\cite{Mikaya82,Janousek79} 
%
  In the flat junction, the benzene molecule was placed 2.7{\AA} above the copper plane with the STM tip 5{\AA} above an orbital $a$.  The tunneling-width matrices for the copper plane and STM tip were set to $\Gamma^{\rm II}_{nm}$=1.3eV and $\Gamma^{\rm I}_{nm}$=$\delta_{na}\delta_{ma}$0.1eV, respectively.  In the para geometry ({cf.}\ top of Fig.\ \ref{fig:benzene_flat_vs_para}), the copper surface and STM leads were arranged opposite one another 3{\AA} away from their closest respective molecular orbitals with a tunneling-width matrix element of 1.3eV.  In both junctions, the voltage was applied symmetrically.

In the previous section, we investigated the dependence of the transmission function on the lead-molecule contact geometry.  From Eq.\ (\ref{eq:current}), it is clear that this dependence directly affects the non-linear current passing through the junction.  As shown in Fig.\ (\ref{fig:flat_benzene_IV}), the near total destruction of Coulomb blockade peaks in the flat configuration gives a nearly flat I-V response, with a nearly two orders-of-magnitude reduction in the current step height as compared with the para junction.  Such quantum interference effects would also affect the apparent height\cite{Weiss93} of adsorbed molecules in STM images.

\section{Conclusions}
\label{sec:conclusions}

In summary, we find that quantum interferences can effectively destroy Coulomb blockade peaks.  The transmission probability of an electron tunneling 
through a flat junction, in which a single-channel lead couples to all $n$ relevant atomic orbitals of a molecule with $n$-fold spatial symmetry, 
is determined by considering the coherent superposition of transmission amplitudes from all possible junction connectivities.  
By virtue of the nodal structure of the many-body wavefunction, these amplitudes exactly cancel at many of the Coulomb blockade resonances, 
completely suppressing the Coulomb blockade peaks but leaving the charging steps unaffected.  

Larger molecules with $n$-fold spatial symmetry, such as [18]-annulene, 
possesses a smaller charging energy than benzene or ethylene, and may therefore be preferred experimentally for Coulomb blockade studies.
We predict that all but the first two transmission resonances in the $\pi$-electron band will be completely suppressed for a flat [18]-annulene junction, 
and that the tails of the
transmission resonances will be even more strongly suppressed than for benzene.

Single-molecule junctions offer a novel and interesting physical system in which to investigate the effects of spatial symmetry and quantum-interference on transport.  Since tunneling transport is central to scanning probe imaging, these interference effects are likely to be important in interpreting images of $n$-fold symmetric molecules adsorbed on conducting surfaces.

PJ acknowledges the support of the NSF under grant No. DMR-0706319.  CAS thanks the Institute of Nuclear Theory at the University of Washington for its hospitality and the Department of Energy for partial support of this work.


%

\end{document}